\documentclass[letter]{aa}
\usepackage{natbib}
\usepackage{graphicx}
\usepackage{txfonts}
\usepackage{comment}
\usepackage{color}
\usepackage{booktabs}
\usepackage[breaklinks,colorlinks=true,citecolor=blue,linkcolor=magenta,urlcolor=blue]{hyperref}

\usepackage{caption}
\usepackage{subfigure}%
\usepackage{rotating}%

\begin{document} 

 \title{Efficient dust radial drift around young intermediate-mass stars}

   \author{
   Paola Pinilla\inst{1,2}, Antonio Garufi\inst{3}, and Matías Gárate\inst{1}.
   }
   \institute{Max-Planck-Institut f\"{u}r Astronomie, K\"{o}nigstuhl 17, 69117, Heidelberg, Germany, \email{pinilla@mpia.de}
   \and Mullard Space Science Laboratory, University College London, Holmbury St Mary, Dorking, Surrey RH5 6NT, UK.
   \and INAF, Osservatorio Astrofisico di Arcetri, Largo Enrico Fermi 5, I-50125 Firenze, Italy.
   }
   \date{}
   
  \abstract{
  The radial velocities and direct imaging observations of exoplanets have suggested that the frequency of giant planets may decrease for intermediate-mass stars ($2.5-8\,M_\odot$). The key mechanism that could hinder their formation remains unclear. From a theoretical point of view, planet formation around intermediate-mass stars may take place on longer timescales, which  -- coupled with fast migration and efficient photoevaporation -- may prevent planetary formation in these environments. In this letter, we investigate the temporal evolution of the radial drift for dust particles in disks when stellar evolution is taken into account. We demonstrate that the particle drift velocity around intermediate-mass stars sharply increases after 1$-$2 Myr, potentially forming a difficult barrier to overcome in the first steps of planet formation. This high
  radial drift could explain the lack of disk detections around intermediate-mass stars older than 3$-$4 Myr, as opposed to low-mass stars ($<2.5\,M_\odot$), where the drift may not be the most impactful factor for the disk evolution. Future high-resolution images of these disks can help us to explain why  planets around intermediate-mass stars may be rare. In addition, we can explore  whether the role of efficient dust radial drift does in fact hinder planet formation around intermediate-mass stars -- or otherwise.}


   \titlerunning{Drift in disks around IMS}
   \authorrunning{Pinilla, Garufi \& Gárate}
   \maketitle

%
%________________________________________________________________
%%%%%%%%%%
\section{Introduction}                  \label{sect:intro}
%%%%%%%%%%

The high occurrence of exoplanets discovered around solar-mass stars indicates that planetary formation around low-mass stars ($<2.5\,M_\odot$) is an efficient process in protoplanetary disks \citep[e.g.,][]{win2015}. Radial velocities and direct imaging observations show that the frequency of giant planets increases as a function of stellar mass, up to a mass of $\sim2.5\,M_\odot$ \citep{johnson2010, reffert2015}. For more massive stars, radial velocity studies show that this frequency quickly decreases \citep{reffert2015}. The detection of planets by radial velocity is sensitive to planets close to the star, so there is a possibility that the decrease in detected planets around stars more massive than $2.5\,M_\odot$ might be attributed to planets forming farther from their parental disk. To test whether this is the case, current direct imaging surveys \citep[such as the BEAST survey using the Spectro-Polarimetric High-Contrast Exoplanet Research at the Very Large Telescope;][]{janson2021} are currently ongoing. In particular, two giant planets in wide orbits in two different systems were recently directly imaged in this survey \citep{janson2021_nature, squicciarini2022}.

\begin{figure} 
    \centering
    \includegraphics[width=9.0cm]{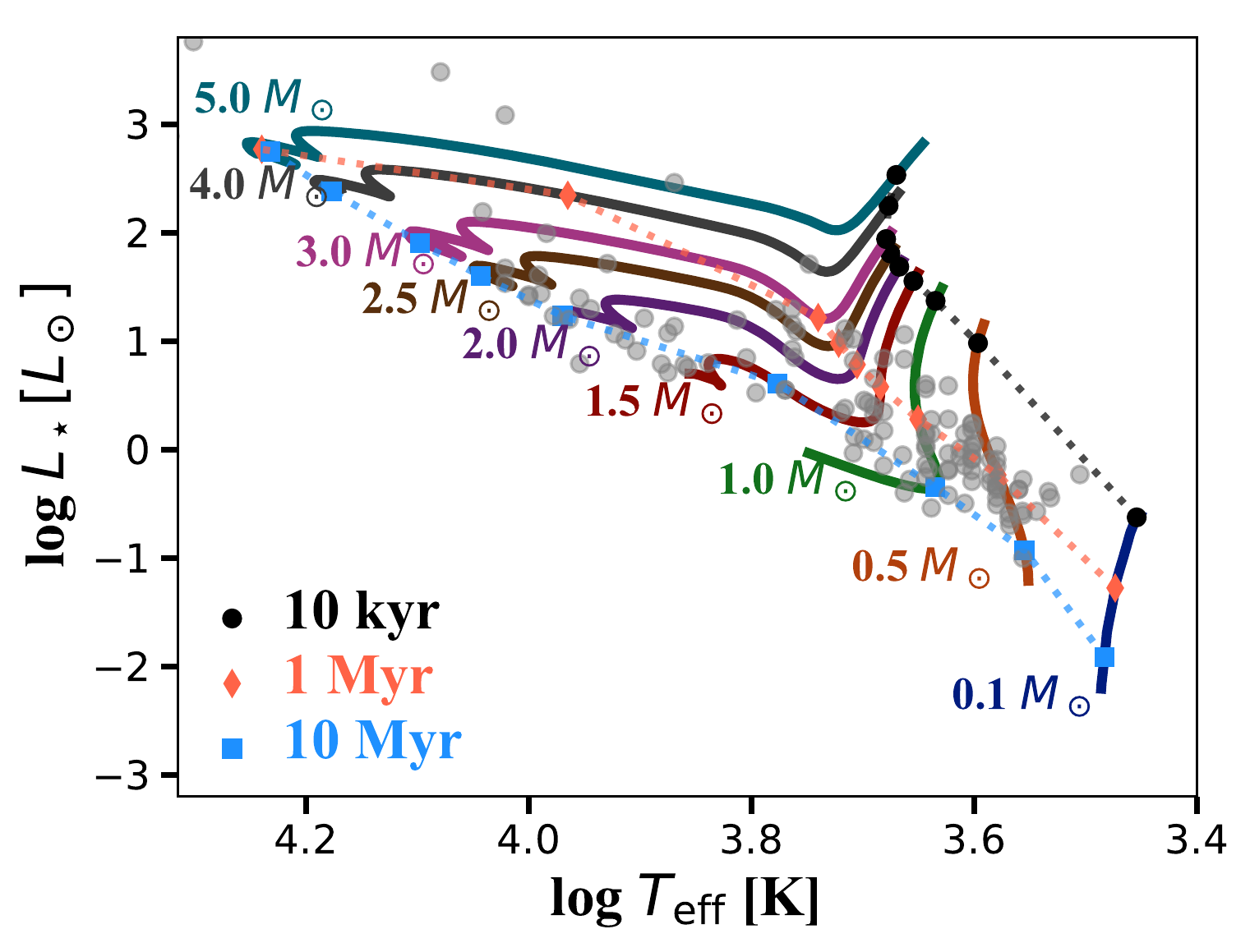}
    \caption{Evolutionary tracks from \cite{dotter2008} up to 30\,Myr of 0.1-5\,$M_\odot$ stars in the H-R diagram, with three different isochronic lines. Points correspond to targets imaged at high angular resolution ($\leq$ 0.1'') in the near-infrared or (sub-)millimeter regimes.}
    \label{fig:pms}
\end{figure}

Intermediate-mass stars (IMS, $2.5-8\,M_\odot$) on the main sequence have spectral types spanning from B1 to B9. We choose 2.5$\,M_\odot$ as the lower limit of this range, because it roughly corresponds to a spectral type of B9 on the main sequence, and thus marks an approximate transition between the A and B spectral type ranges \citep{pecaut2013}. During the pre-main-sequence (PMS) phase when planetary formation is thought to occur, the effective temperature of the IMS is low. Depending on the mass, an IMS spends from a fraction of a Myr to a few Myr of its initial life with spectral type later than F0 being therefore defined as an intermediate-mass T Tau star \citep[IMTTS, see e.g.,][]{Calvet2004}. It later shifts to an earlier spectral type and appears as an Herbig Ae/Be star \citep{Waters1998}. Their luminosity is also significantly modified throughout the PMS phase, initially decreasing and subsequently increasing until the star reaches the main sequence (Fig.~\ref{fig:pms}). 

The apparent sharp decrease in the number of giant exoplanets around IMS opens up questions about the physical conditions of protoplanetary disks around Herbig Ae/Be and IMTTSs and what effects could prevent (giant) planetary formation in these environments.  Infrared observations have demonstrated that the disk lifetime around the IMS is shorter than that of low-mass stars \citep{carpenter2006, ribas2015, luhman2022}. The inner parts ($<20\,R_\star$) of the disks around the IMS are thought to evolve due to viscous decretion, where the star rotating near breakup velocity at the equator leads to episodic or continuous mass ejection from
the stellar equator. As a result, outflowing, ionized, purely gaseous disks spreads viscously into a disk and reaccretes \citep[e.g.,][]{lee1991, bjorkman2005, rivinius2013}. At larger radii, it is still unclear what dominates the evolution of the disks around an IMS and, thus, observations that are sensitive to regions much larger than 20 stellar radii are needed to understand their evolution \citep[e.g.,][]{klement2017}.

From a theoretical point of view, planetary formation may also decrease at masses larger than $2-3\,M_\odot$ \citep[e.g.,][]{Ida2008, kennedy2008}. This is because when the mass of the star increases, the snow line is located further out, where the growth timescales for planetary formation by planetesimal accretion are longer than the migration timescales, while simultaneously effective photoevaporation of the disk driven by energetic stellar photons is more effective in these disks \citep[e.g.,][]{adams2004}. This combination of slow growth to planets, fast migration, and efficient photoevaporation may prevent planetary formation around IMS. However, planets may still form around IMS as around Sun-like stars by pebble (millimeter- and centimeter-sized particles) accretion, which occurs faster than planetesimal accretion \citep{ormel2010, Lambrechts2012}.

In either scenario (pebble or planetesimal accretion), the very first step of planet formation is the growth from micron-sized particles from the interstellar medium to pebbles. In this process, dust particles are expected to drift towards the regions of high pressure because of the sub-Keplerian rotation of the gas \citep[e.g.][]{Weidenschilling1977}. The drift velocity of the particles depends on how much the gas velocity differs from Keplerian rotation. This difference from Keplerian rotation changes across the stellar masses and age \citep[see][for the case of disks around very low mass stars]{pinilla2013}. The dust drift velocity is such that $v_{\rm{drift}}\propto L_\star^{1/4}/\sqrt{M_\star}$. The dependency of the stellar mass dominates for low mass stars and, as a consequence, the radial drift velocities are higher for particles in disks around very-low-mass stars ($\lesssim0.1\,M_\odot$) and brown dwarfs  ($\lesssim$ 5\%\,$M_\odot$) than in disks around Sun-like stars. This prediction has been recently supported by observations from Atacama Large Millimeter/submillimeter Array (ALMA) of these objects, where the gas disk extension traced by $^{12}$CO is much larger than the dust continuum emission in the absence of substructures that may act as pressure bumps  \citep[the disk around CIDA\,7, ][]{kurtovic2021}.

Because the luminosity of an IMS significantly increases with time after a fraction of million-of-years, the luminosity term dominates the drift velocity of dust particles ($v_{\rm{drift}}\propto L_\star^{1/4}/\sqrt{M_\star}$) for disks around IMS. As a result of the increase in luminosity, the gas velocity can soon become more sub-Keplerian around these stars, inducing a very efficient drift of dust particles before the first embryos of planets can form. 

In this letter, we investigate the radial drift of dust particles in disks around IMS when the stellar luminosity and temperature vary over time (Sect.~\ref{sect:drift}), similarly to the work of \cite{appelgren2020} for T-Tauri stars. In addition, in this section we compare our results with the fraction of detected protoplanetary disks at high angular resolution around a large range of stellar masses and ages. We discuss the implications of our results in Sect.~\ref{sect:discussion} and we summarize our findings in Sect.~\ref{sect:conclusion}. 

%%%%%%%%%%
\section{Radial drift of dust particles in disks around an IMS}                    \label{sect:drift}
%%%%%%%%%%
\begin{figure*} 
    \centering
    \tabcolsep=0.05cm 
    \begin{tabular}{cc}   
        \includegraphics[width=9cm]{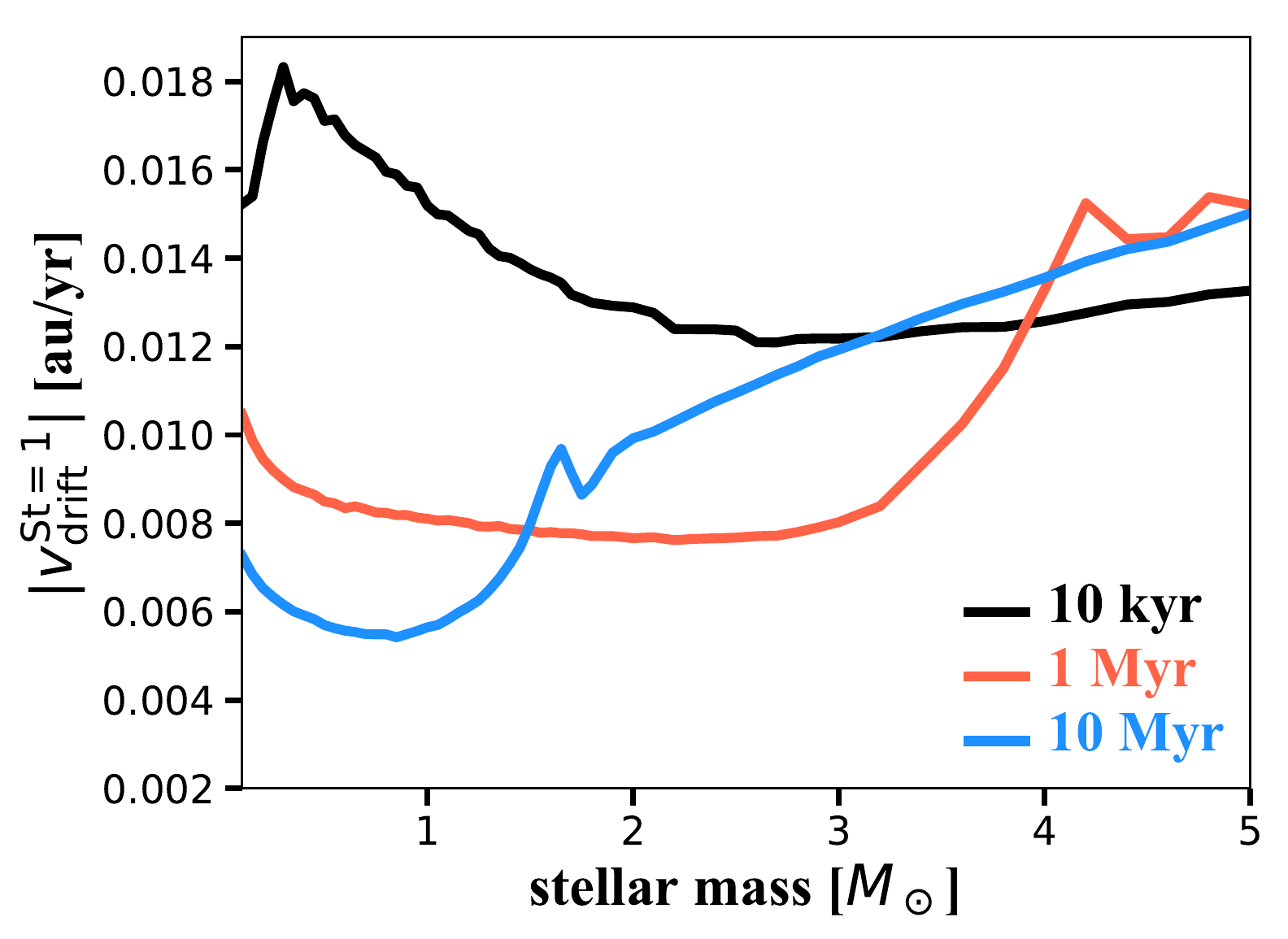}&
        \includegraphics[width=9cm]{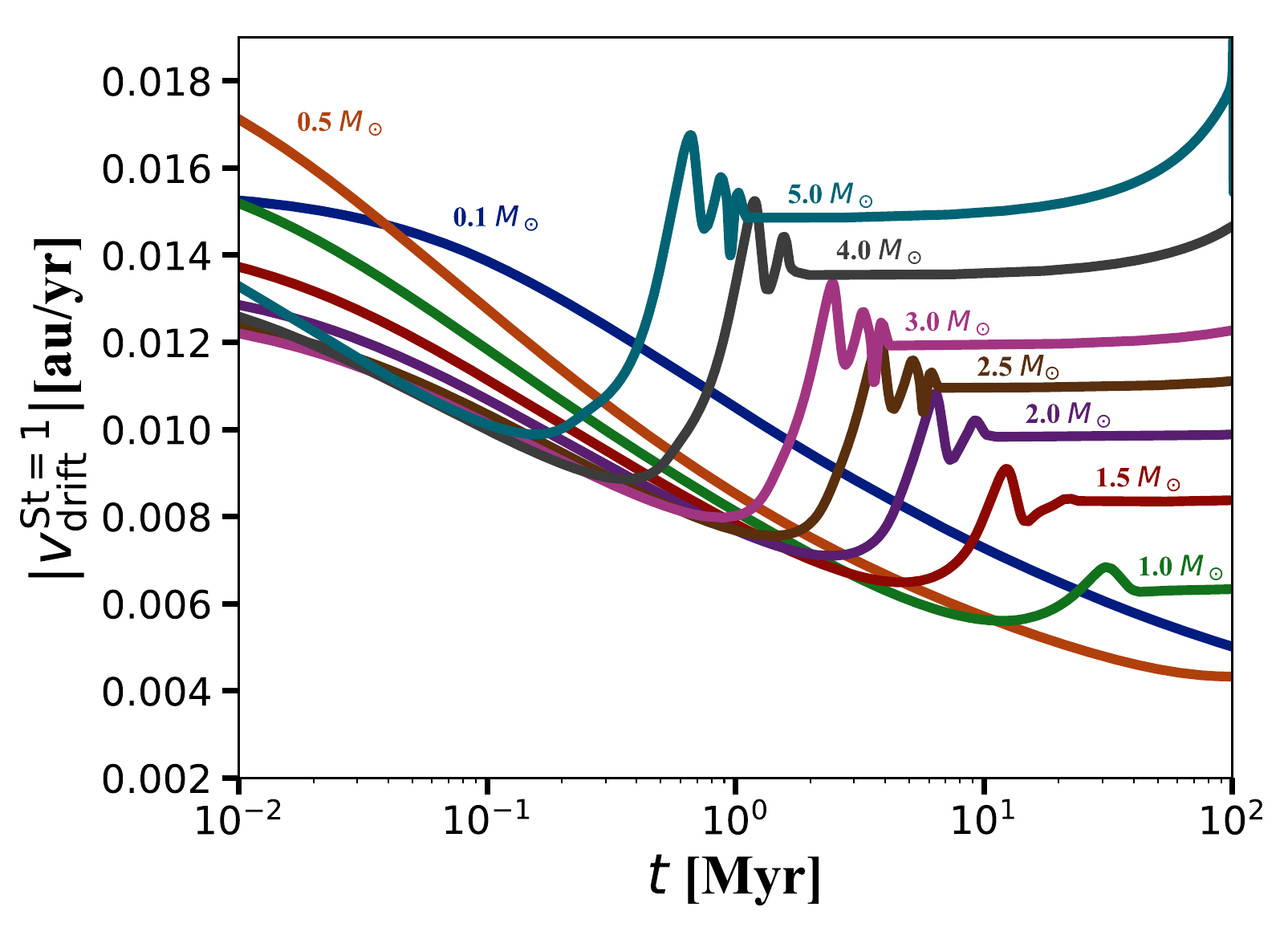}
    \end{tabular}
    \caption{Evolution of the maximum dust drift velocity when considering stellar evolution. Left panel: Absolute value of the dust drift velocities at the midplane as a function of stellar mass for particles with St=1, and at three different times of stellar evolution following the evolutionary tracks from Fig~\ref{fig:pms}. Right panel: Absolute value as a function of time for the drift velocity of dust particles with St=1 in the disk midplane around stars with different masses.}
    \label{fig:drift}
\end{figure*}

The total drift velocity of dust particles in protoplanetary disks is given by \citep{Weidenschilling1977}:
\begin{equation}
        v_{\mathrm{drift}}(\mathrm{St}, M_\star, L_\star)=\frac{1}{\textrm{St}^{-1}+\textrm{St}} \frac{\partial_r P (M_\star, L_\star)}{\rho(M_\star, L_\star) \Omega(M_\star)}. 
        \label{eq:vdrift} 
\end{equation}

The dependency of $v_{\mathrm{drift}}$ on the stellar mass and luminosity is as follows. First, the Stokes number denoted by St quantifies the coupling of the particle to the gas, and in the midplane, this is defined as

\begin{equation}
        \textrm{St}(M_\star)=\frac{a\rho_s}{\Sigma_g (M_\star)}\frac{\pi}{2},
        \label{eq:stokes}
\end{equation}

\noindent where $a$ is the particle size, $\rho_s$ is the solid density of the dust grains, and $\Sigma_g$ is the gas surface density that depends on the stellar mass under the assumption that the disk mass scales with the stellar mass. For this purpose, we assume $M_{\rm{disk}}=0.05\,M_\star$ and a power law for the gas surface density: $\Sigma_g\propto r^{-1}$. 

Second, in Eq.~\ref{eq:vdrift}, we set  $\rho$($M_\star, L_\star$) as the disk gas volume density, which in hydrostatic balance and in a vertically isothermal disk is: 

\begin{equation}
        \rho=\rho_0(M_\star, L_\star)\exp\left(\frac{-z^2}{2h^2(M_\star, L_\star)}\right)
        \label{eq:rho}
,\end{equation}

\noindent where $\rho_0(M_\star, L_\star)$ is the midplane density  and $h(M_\star, L_\star)$ is the disk scale height, which are defined as:

\begin{equation}
        \rho_0=\frac{\Sigma_g(M_\star)}{\sqrt{2\pi}h(M_\star, L_\star)} \qquad \textrm{and} \qquad h(M_\star, L_\star)=\frac{c_s(L_\star)}{\Omega(M_\star)},
        \label{eq:rho_0}
\end{equation}

\noindent where the sound speed ($c_s$) is given by:

\begin{equation}
   c_s^2= \frac{\sigma_{\textrm{SB}} T_{\textrm{disk}} (L_\star)}{\mu m_p},
        \label{eq:cs}
\end{equation}

\noindent with $\sigma_{\rm{SB}}$, $\mu,$ and $m_p$ the Stefan-Boltzmann constant, the mean molecular mass, and  the proton mass, respectively. We assume the disk temperature ($T_{\rm{disk}}$) of a passive disk as \citep{kenyon1987}:

\begin{equation}
        T(r, L_\star)=T_\star\left(\frac{R_\star}{r}\right)^{1/2} \phi_{\rm{inc}}^{1/4} =\left(\frac{L_\star \phi_{\rm{inc}}}{4\pi \sigma_{\textrm{SB}} r^2}\right)^{\frac{1}{4}}
  \label{eq:temp}
,\end{equation}

\noindent where $R_\star$ and $T_\star$ are the stellar radius and temperature, respectively, and  $\phi_{\rm{inc}}$ is the incident angle, which we assume to be 0.05. 

Finally, the isothermal pressure is given by:
\begin{equation}
    P = \rho(M_\star, L_\star) c_s^2(L_\star).
    \label{eq:pressure}
\end{equation}

Considering these dependencies on stellar mass and luminosity, \cite{pinilla2013} demonstrated that $v_{\mathrm{drift}}\propto L_\star^{1/4}/\sqrt{M_\star}$. The left panel of Fig.~\ref{fig:drift} shows the absolute value of the dust drift velocities (Eq.~\ref{eq:vdrift}) at the midplane as a function of stellar mass for particles with St=1 (which corresponds to the highest radial drift velocities) and at three different times of stellar evolution following the evolutionary tracks from Fig~\ref{fig:pms}. At early times (10\,kyr), dust particles in disks around stars with $M_\star<1\,M_\odot$ experience the highest radial drift, while at later times (1 and 10\,Myr), the drift velocities are high for very-low-mass stars ($\sim$0.1\,$M_\odot$) and increase for an IMS ($>2.5\,M_\odot$). It is important to note that the  radial drift for particles around stars even less massive than $\sim$0.1\,$M_\odot$, as well as brown dwarfs or even planetary objects, is expected to continue increasing, as shown by \cite{pinilla2013} and \cite{zhu2018}.

The right panel of  Fig.~\ref{fig:drift} shows the absolute value of the dust drift velocity in the disk midplane as a function of time for particles with St=1 and for different stellar masses. Because we calculate the radial drift velocity for a given St rather than a given grain size, these results are independent on the radial location where these velocities are calculated (we note that for the above assumptions $\partial_r P/(\rho_0 \Omega$) is constant with the radius).  For 0.1, 0.5, and 1\,$M_\odot$ the drift velocities decrease with time, while for stellar masses $M_\star>1.5\,M_\odot$ an increase in the drift velocities occurs within the typical protoplanetary disk lifetime (up to 10\,Myr). This increase is sharper, higher, and occurs earlier for more massive stars. The same trend is expected for lower values of St, but with lower drift velocities. Dust particles in protoplanetary disks may grow to sizes near St=1 depending on different dust and disk properties, such as disk turbulence and the fragmentation velocity of the particles \citep{pinilla2021}.  

Although the difference of drift velocities between different stellar masses seems moderate, it can significantly change one of the first steps of planet formation, which is the formation of pebbles, as well as their long-term retention by trapping in pressure bumps. In the case of very-low mass stars, where the drift velocities are \emph{only} a factor of two higher (at most) than around Sun-like stars (also seen in Fig.\,\ref{fig:drift}), the amplitude of the pressure gradient of the bumps needs to be higher in order to stop the drift of the grains and the effect of turbulent diffusion inside these traps \citep{pinilla2013}. Therefore, in disks around IMS, a higher pressure gradient would be required to stop the drift of grains.

Figure~\ref{fig:obs_theory} compares the age and mass of stars imaged with high angular resolution ($\leq$0.1'') either in the near-infrared  or at (sub-)millimeter wavelengths  with the times when radial drift velocities reach their maximum value for different stellar masses in the right panel of Fig.~\ref{fig:drift}. The data-points from near-infrared observations are listed in \citet{garufi2018, garufi2020, garufi2022} and \citet{laws2020}. Sources with ALMA observations can be found in \citet{long2018}, \citet{andrews2018}, \citet{cieza2019}, \citet{ansdell2020},  \citet{villenave2020}, \citet{francis2020}, \citet{kurtovic2021}, and \citet{stapper2021}. The stellar masses and age of Fig.\,\ref{fig:obs_theory} are calculated for all sources where the star is visible following the method outlined by \citet{garufi2018} and based on the latest \textit{Gaia} data release \citep{gaia2021}.

On one hand, the desert of observational data in the lower right of Fig.~\ref{fig:obs_theory} is due to the faint luminosity of the host star that does not allow us to carry out optical/NIR high-contrast imaging. In fact, these observations are aided by an adaptive optics system that requires a relatively bright star to operate \citep[see][]{benisty2022}. On the other hand, the upper desert is most likely an actual paucity of disks still present around the host star. The similarily between the boundary of this desert and the time where drift velocities are maximized is discussed in Sect.\,\ref{sect:discussion}.

%%%%%%%%%%
\section{Discussion}                    \label{sect:discussion}
%%%%%%%%%%

As protoplanetary disks evolve, so do the host stars. A main shortcoming of most of the simulations studying the disk evolution is to ignore the impact of stellar evolution on the disk. By including the stellar luminosity from evolutionary tracks, we show that the dust radial drift in disks around low- and intermediate-mass stars has a different behavior. In low-mass stars, the drift is more pronounced at the very early stage ($\ll$1 Myr, left panel in Fig.~\ref{fig:drift}) and, over time, the drift velocities increase sharply for more massive stars (right panel in Fig.~\ref{fig:drift}). Radial drift may become a more difficult  barrier to overcome in planet formation in  disks older than 1 Myr around an IMS. This is because the first steps of planet formation feature growth from (sub-)micron-sized particles to pebbles (particles with St near unity). Once dust particles grow to these sizes,  gas-giant cores or terrestrial planets can quickly form thanks to streaming instability and pebble accretion \citep[e.g.,][]{Lambrechts2012}. It is possible that due to the very fast radial drift of dust particles in disks around IMS, grains cannot grow to pebble sizes (and furthermore to planetesimals) due to destructive collisions with high drift velocities; another possibility is simply that they are lost towards the star before any pressure bump can form in the disk to stop their drift. This finding may provide an explanation for the paucity of planets detected around these stars that is alternative (or concurrent) to other processes such as slow growth to planets, fast planet migration, and efficient photoevaporation. 

\begin{figure} 
    \centering
    \includegraphics[width=9.0cm]{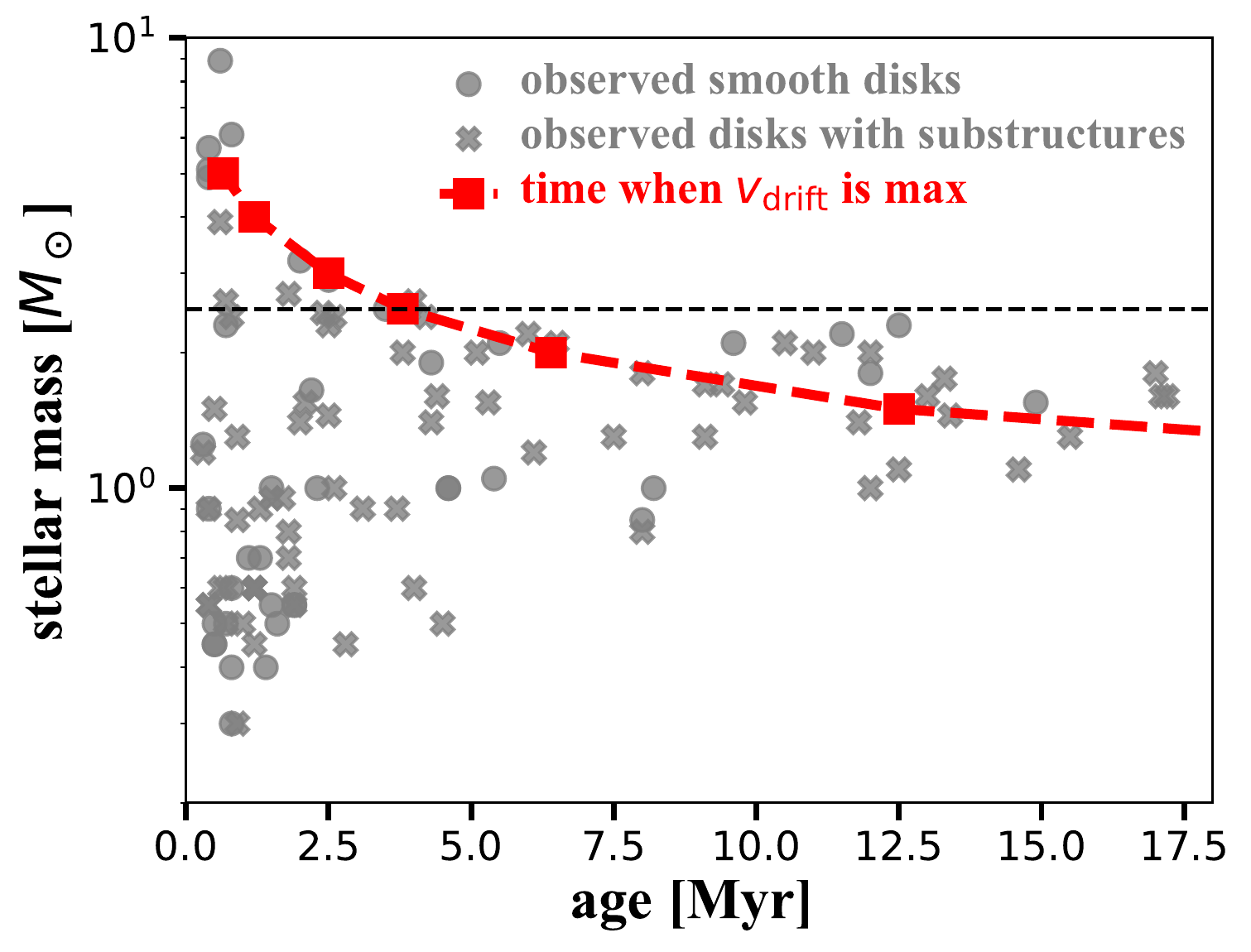}
    \caption{Age and mass of stars imaged at high angular resolution ($\leq$0.1'') compared with the times where the radial drift velocities of disk particles reach their maximum value for different stellar masses (right panel in Fig.~\ref{fig:drift}) and following the evolutionary tracks shown in Fig~\ref{fig:pms}. Circles are disks that are identified as smooth disks, while crosses are disks where substructures have been detected. The black dashed-line is at $2.5\,M_\odot$.}
    \label{fig:obs_theory}
\end{figure}

The dichotomous evolution of the dust radial drift around low mass stars and IMS must therefore have an observable effect on the disk morphology around these two classes of stars. The abrupt change in the temporal evolution of the drift around a 1\,M$_\odot$ star (decreasing) and $>2.5\,M_\odot$ stars (sharply increasing)  may have an immediately observable imprint on the disk demography: several solar-mass stars as old as 10 Myr are known to still host a relatively massive planet-forming disk while no disk has ever been imaged around any IMS older than 3$-4$ Myr (Fig.~\ref{fig:obs_theory}). This suggests that the rapid increase in luminosity experienced by an IMS after 1 Myr of life accelerates the disk evolution to the point that the whole disk is dissipated in a fraction of Myr.

As for the disk morphology around young IMS, not much is known. In fact, at present, it is only a few stars with spectral types earlier than B9 have been imaged with high angular resolution \citep[e.g., MWC297 and HD87643,][]{stapper2021, laws2020}. Among the few constraints from unresolved observations available is their low dust mass compared to low-mass stars (a few M$_\oplus$) and loose upper limits on the extent \citep[often as much as a few hundred au,][]{alonso2009, stapper2021}.  
It is therefore pivotal to carry out a survey of these targets at a high angular resolution and sensitivity that could shed light on their disk morphology. A fundamental question that arises is whether these disks are similar in mass and extent to the coeval counterparts around low-mass stars or whether they are already smaller when the onset of the efficient radial drift and photoevaporation act to dissipate them. Our findings (shown in Fig.~\ref{fig:obs_theory}) suggest that radial drift is an impacting factor driving the disk evolution, limiting the disk detection around an IMS, while for lower mass stars, the effect of drift is less clear; this is potentially due to the presence of pressure bumps that could form early in these disks, helping to retain dust particles. In fact, most of the disks around low-mass stars ($<2.5\,M_\odot$) do show substructures (as shown in Fig.~\ref{fig:obs_theory}). However, no obvious trend is seen between the presence or absence of substructures in the disks around low-mass-stars, as illustrated by the red line from the models in Fig.~\ref{fig:obs_theory}.

Given the radial drift of dust particles  is more efficient around the IMS may have an important effect on the stellar photosphere- and inner disk composition. \cite{kama2015}  proposed that the depletion of heavy elements in the photospheres of Herbig Ae/Be stars hosting disks could come from pressure bumps (possibly from giant planets forming in these disks) blocking the dust accretion in the outer disk.  If planet formation is hindered by very efficient radial drift around an IMS, there should have higher abundances of refractory elements involved than low-mass stars; alternatively, another process must be responsible for the depletion of metals if no planets or sub-structures are found in future high resolution observations. \cite{banzatti2020} also suggests that drifting pebbles from the outer parts of the disks enrich the inner disk with water, while the water abundance is lower in disks where the pebbles are trapped. For IMS disks, the inner disk should therefore be more water rich than their low-mass stars, a hypothesis that could be tested with incoming data from the James Webb Space Telescope (JWST).

%%%%%%%%%%%%
\section{Conclusion} \label{sect:conclusion}
%%%%%%%%%%%%%%

Radial velocities and direct imaging observations of exoplanets suggest that the frequency of giant planets decreases for IMS $2.5-8\,M_\odot$ and it is unclear what key mechanism is responsible for hindering their formation around these stars. In this letter, we demonstrate that because the luminosity of IMS significantly increases over time after a fraction of a million years, the radial drift velocity of dust particles increases as well, potentially becoming an important barrier for planet formation. While for low-mass stars ($<2.5\,M_\odot$), radial drift is not necessarily the dominant factor of evolution, either because pressure bumps (possibly from planets forming in these disks) are present or because other mechanisms drive the main disk and dust evolution. For the IMS, the efficient radial drift velocity could explain the lack of disk detection around these objects at ages older than 3$-$4 Myr, but other physical mechanisms may play an important role as well, such as photoevaporation or viscous decretion. It is crucial to reveal the disk morphology around an IMS based on high angular-resolution and high-sensitivity observations at multiple wavelengths in order to alleviate one major shortcoming of the current observational census of young stars. In turn, these observations will help us explain why planets around IMS may be rare, in addition to shedding light on the role of radial drift as a barrier to planet formation around any class of stars.

%%%%%%%%%%%%
\section{Acknowledgements}
We thank the referee for his/her/their report that helped us to clarify different aspects of this manuscript. We are thankful to the organizers of the online conference "Five years after HL Tau: a new era in planet formation", where this idea emerged. We are also thankful with Carsten Dominik, Myriam Benisty, and the Genesis of Planets group at MPIA for insightful discussion. Authors acknowledge support provided by the Alexander von Humboldt Foundation in the framework of the Sofja Kovalevskaja Award endowed by the Federal Ministry of Education and Research. 
  
%%%%%%%%%%%%%%

\bibliographystyle{aa} % style aa.bst
\bibliography{ref}
\end{document}